\documentstyle[aaspp4,12pt]{article}
\received{}
\accepted{}
\def\l{$\lambda$}
\def\ne{n$_{\rm e}$\/}

\def\REF{\par\noindent\hangindent 20pt}
\def\msol{M$_\odot$\/}

\def\rg{R$_{\rm g}$\/}
\def\ltsima{$\; \buildrel < \over \sim \;$}
\def\simlt{\lower.5ex\hbox{\ltsima}}            
\def\gtsima{$\; \buildrel > \over \sim \;$}

\def\simgt{\lower.5ex\hbox{\gtsima}}            
 
\def\a{$\alpha$}
\def\hiha{{\sc Hi~H}$\alpha$\/}
\def\Ka{{\sc K}$\alpha$\/}
\def\feka{{\sc Fe~K}$\alpha$\/}
\def\ha{{\sc H}$\alpha$}
\def\civ{{\sc{Civ}}$\lambda$1549\/}

\def\cm3{cm$^{-3}$\/}

\def\mgii{{\sc{Mgii}}$\lambda$2800\/}

\def\o4363{{\sc{[Oiii]}}$\lambda$4363\/}

\def\feii{{\sc{Feii}}$_{\rm opt}$\/}

\def\kms{km~s$^{-1}$}
\begin{document}
\title{\sc On the Origin of Broad Fe \Ka\ \&\ Hi H$\alpha$ Lines in 
AGN\footnote{Based in part on observations collected at ESO, La Silla}}
\author{J. W.\ Sulentic\altaffilmark{2}, P. Marziani\altaffilmark{3}$^,$\altaffilmark{4}, 
T.\ Zwitter\altaffilmark{5}, M.\ Calvani\altaffilmark{3} and D.\ 
Dultzin-Hacyan\altaffilmark{4}}

\altaffiltext{2}{Department of Physics \&\ Astronomy, University of Alabama, Tuscaloosa, 
AL 35487, USA. E-mail: giacomo@merlot.astr.ua.edu}

\altaffiltext{3}{Osservatorio Astronomico, Padova, Italy}

\altaffiltext{4}{Instituto de Astronomia, UNAM, Mexico City, DF 04510,  Mexico}

\altaffiltext{5}{Department of Physics, University of Ljubljana, Ljubljana, Slovenia}

\begin{abstract}

We examine the properties of the {\sc Fe} emission lines that arise near 6.4
keV in the ASCA spectra of AGN. Our emphasis is on the Seyfert 1
galaxies where broad and apparently complex \feka\  emission is observed.
We consider various  origins for the line but focus on the pros and
cons for line emitting accretion disk models. We develop a simple
model of an illuminated disk capable of producing both X-ray and optical lines
from a disk. The model is able to reproduce the observed \feka\
FWHM ratio as well as the radii of maximum emissivity implied by the
profile redshifts.  The overall profile shapes however do not fit well
the predictions of our disk illumination model nor do we derive always consistent 
disk inclinations for the two lines. We conclude that the evidence for 
and against an accretion disk origin for the \feka\ emission is equal at 
best.  The bulk of the data requires a very disparate set of line fits 
which shed little light on a coherent physical model. We briefly consider 
alternatives to disk emission models and show that a simple bicone model 
can reproduce the {\sc Fe} line profiles equally well.

\end{abstract}

\keywords{Galaxies:  Active -- Galaxies:  Kinematics \&\ Dynamics --
Galaxies: Nuclei -- Galaxies:  Seyferts -- Galaxies: Emission Lines -- Galaxies:
X--Rays -- Line:  Formation -- Line:  Profiles}

\section{Introduction}

The idea that a significant fraction of the broad line emission in 
Active Galactic Nuclei (AGN)
comes directly from an accretion disk is attractive for several reasons
beyond the obvious one of proving the black hole paradigm.  The disk
provides a high density medium where the integrated intensities of the
optical {\sc Feii} lines can be accounted for. An illuminated disk can also
more easily produce the strong Balmer lines (i.e. the L$\alpha$/H$\beta$
problem: see Collin-Souffrin \&\ Dumont 1990). Apparent density and kinematic
differences between high and low ionization lines (Gaskell 1982; 
Sulentic 1989) can also be explained if the latter arise from a disk. 
Naturally there was considerable excitement when the double-peaked Balmer 
lines in Arp102B were discovered (Chen, Halpern and Filippenko 1989). 
The line profiles in that AGN could be well-fit by the predictions of a 
model for line emission from a relativistic Keplerian disk (Chen and 
Halpern 1989). The fit to Arp102B was quite good but the statistical 
implications were not.  The best disk model fits to lines in Arp102B 
imply an intermediate viewing angle and that leads to the expectation 
of many sources with double-peaked emission lines. A  search for more 
double peaked and peculiar profiles among radio-loud
AGN revealed only a handful that could be reasonably well fit by disk
models (Eracleous and Halpern 1994).

AGN with optical line profiles like Arp102B are apparently quite rare. 
They also appear to be almost uniquely radio-loud sources. Either most
radio-loud disk emitters (and all radio-quiet ones)  produce emission
at larger radii where double peaked lines are not expected or the disk
contribution to the broad line emission is small or negligible. The
unsatisfactory alternative would be that we are viewing virtually all
AGN aligned with the disk axis near pole-on. A comparison between line profile
shifts/asymmetries with disk model predictions (Sulentic et al.\ 1990)
suggests that at least part of the line emission in most quasars must arise
from another source. This conclusion is based on the
large number of line profiles that show blue shifts and/or blue
asymmetries (implying that the red side of the profile is brighter) in
conflict with disk model predictions.  Even the most promising
double-peaked sources (e.g.\  Arp102B, 3C390.3) require fine tuning of
the disk models (see e.g.\ Gaskell 1996; Zheng et al.\ 1991) because the
profile peaks appear to vary out of phase. A new double peaked source
arose in Pictor A with first a red, and then a blue, peak developing
over a five year period (Sulentic et al.\ 1995a). Such variations and
the overall rarity of double-peaked profiles appear to be more easily
explained with models invoking a predominance of radial motion. Such
models also have the advantage that they can directly account for the
rarity of peculiar line profiles because such profiles would arise at
an extremum in the orientation/kinematic domain (Sulentic  et al.\ 1995a).

A possible kinematic difference between high and low  
ionization lines (HIL and LIL; HIL include \civ\ and {\sc Heii} lines; LIL are
Balmer lines, \feii\ lines, and \mgii) was one of the reasons that models for a two zone 
broad line region (disk + something else) became necessary and popular. 
The evidence revolves around a possible systematic blueshift of the  
HIL with respect to the LIL (the LIL showing centroid velocities 
similar to the source rest frame).  The most peculiar line profiles 
were found in the radio-loud AGN population and subsequent searches 
for disk candidates focused there (Eracleous and Halpern 1994).  
Ironically the largest sample yet studied for line profile shifts 
indicates that it is only the {\it radio-quiet} AGN that show a 
systematic HIL blueshift with respect to the  LIL (Sulentic et al.\ 
1995b; Marziani et al.\ 1996).

Recent advances in X-ray spectral resolution  offered by the Advanced
Satellite for Cosmology and Astrophysics (ASCA; Tanaka, Holt, \&\ Inoue 1994)
provide a possible new form of evidence for disk emission in (predominantly)
radio-quiet AGN. We may be observing  the very innermost  regions of the disk
that could be producing an iron K$\alpha$ line by Compton reflection or
photo-ionization (e. g. Tanaka et al. 1995; Fabian et al. 1995; Dabrowski et
al. 1997; Fanton et al. 1997). Strong and broad line  emission is often
observed near 6.4 keV in AGN spectra (Mushotzky, Done, \&\ Pounds 1993;
Mushotzky 1997; Nandra et al. 1997a,b, and references therein).  
Figure 1 illustrates the difficulties that we face in quantifying the X-ray line emission. While
ASCA provides a tremendous advance in X-ray spectroscopy, the resolution
(2-5\% at 6 keV)  and S/N obtainable still lag far behind optical data.   We
show a typical modern observation of the H$\alpha$ line profile for Arp102B
degraded to the resolution and S/N of a typical ASCA spectrum.  Narrow lines
near H$\alpha$ arising from {\sc [NII]}\l\l6548,6583, {\sc [OI]}\l 6300 and
{\sc [SII]}\l\l 6717,6730 were removed from the spectrum before blurring
because the \feka\  emission in AGN has no known analogous contaminants,
albeit \feka\ might  involve a blend of multiple, broad and narrow, components. 
While optical observations can discriminate between a disk model and a  broad
gaussian, ASCA observations cannot (see Fig. 1).

We consider {\sc Fe} line detection statistics for different AGN classes in 
\S\ \ref{xopt}.  In \S\ 3 we consider an accretion disk origin for the \feka\ 
and \hiha\ lines.  In \S\ 3 we model an
illuminated disk (e.g. Dumont \&\ Collin-Souffrin 1986)  to account for both optical and X-ray
emission lines simultaneously.  In \S\ 4 we summarize  problems with the
accretion disk interpretation for \feka\ and  consider an  alternative model
for the X-ray line emission.

\section{X-ray and Optical Observations \label{xopt}}

The energy of the emitted \feka\ depends on ionization stage. The ASCA
resolution allows us to distinguish line emission from three main ranges of
ionization: (1) emission from neutral or weekly ionized iron (``cold iron'' emission; 
ionization stage
$\leq${\sc Fexviii}), as expected from cold, dense gas in an accretion disk; (2) emission from mildly to highly ionized
iron at 6.45 keV $\simlt$ energy $\simlt 6.7$ keV, for  {\sc Fexix}$\leq$ ionization
stage $\leq${\sc Fexxv}; (3) emission  from H-like {\sc Fe} at 6.97 keV ({\sc
Fexxvi}).  The most robust measurement  is the centroid energy of the line,
but even this can become confused when the line is broad. Determination of the
profile width, asymmetry index or equivalent width are more difficult and
model dependent. These parameters, and the derived $\chi^2$\ goodness-of-fit
are quite sensitive to the continuum fit which includes a basic power law
modified by both emission and absorption processes (correlated to the line
itself). 

\subsection{Statistics of Emission lines near 6.4 keV}

Table 1 summarizes the published {\sc Fe} line parameters for most AGN with
detected \feka\  emission in ASCA spectra. The list should be reasonably
complete through 1996. The format for Table 1 is as follows:
Column 1 - abbreviated (2000) position taken from the 7th edition of
the Veron \& Veron catalog; Column 2 - a common name for the source as
used in the ASCA-related literature; Column 3 - AGN type; Column 4 -
optically determined redshift; Columns 5, 6, and 7 - the derived \feka\
line peak centroid energy, profile Gaussian width (keV) and EW (eV),
respectively; Column 8 - reference(s) to the source of the \feka\
parameters given in the three previous columns. Table 1 reflects
the best currently derived line parameters. In some cases a range is given
to reflect a range of reasonable solutions, providing additional
insight into the uncertainties associated with these numbers.
References should be consulted  for details of adopted continuum and
line properties. We focus our discussion on the Seyfert galaxies
because: (1) that is where the largest body of X-ray emission line data
exists and the statistics of \feka\  emission from broad-line radio 
galaxies and quasars are not yet clear and (2) the \feka\ spectra of Sy 1
galaxies sometimes show ``disk-like'' profiles. 

A recent study by Nandra et al.\ (1997a,b) has collected much of the data and
reduced it uniformly. The best model fits to Seyfert 1 spectra always show
{\sc Fe} emission with approximately 80\% of the detections better fit by  a
resolved broad line.  Nandra et al.\ (1997b) find a distribution of profile
(Gaussian fit $\sigma$) widths with: (1) about half of the sample clustered at
0.7$\pm$0.1 keV and (2) the other half of the sample extending continuously from
0.5 keV down to unresolved instrumental width ($\sim$0.1 keV). The profiles frequently
show a narrow component centered  very near to 6.4 keV accompanied by an
asymmetry towards the red. In  several cases the spectra suggest a
multicomponent structure within this   red wing rather than a smooth
extension. In most cases it is impossible  to determine if this multicomponent
structure is real or due to  instrumental effects and low S/N.  The
distribution of line equivalent  widths extends from 50-600 eV with a mean
EW$\sim$160. Typical  uncertainties for the individual estimates cover most of
this range. The  major point of consensus is that we typically observe a broad
component  extending 1-1.5 keV to the red of a narrow peak that is centered
very close  to 6.4 keV. The two simplest models of the line involve: (1) a
single asymmetric feature (\S\ 3) or (2) a combination of two symmetric 
features one narrow/unshifted and another broad/redshifted (\S\ 4). 

The
situation for radio quiet quasars is very confused with many showing no
evidence of rest frame emission near 6.4 keV. At the same time sources from
3C273 to S5 0014+813 (Elvis et al.\ 1994) at z=3.384 show evidence for a
line.  Rest frame centroid energies for these detections range from
6.4-6.9keV.  Few observations for radio-loud sources  have been published and
the situation is not clearly defined. No radio-loud sources as yet show broad
{\sc Fe} profiles like IRAS18325-5926 or Sy 1's, as further discussed in \S\
\ref{stat}.  In 3C390.3 
(Eracleous et al.\  1996) the \ha\ and \Ka\ widths  are similar while for
3C120 \feka\ may show either a single broad line or 3  individual narrower 
features. A recent observation for Pictor A  (Halpern et al.\ 1997) shows no
{\sc Fe} line. Recent claims  (Reynolds 1997) that radio loud sources show
much broader \feka\ profiles  than radio quiets are based upon very broad and
uncertain detections for  3C120 and 3C382. 

Seyfert 2 sources usually show narrow \feka\ emission.  The results from a
large GINGA sample (Smith \& Done 1996) indicate that most lines have a
centroid near 6.4 keV with a subset centered between 6.6-7.0 keV.  In two of
the best studied cases  (Mark 3 and NGC1068) there is evidence for both
neutral and ionized {\sc Fe} lines, possibly components of {\sc Feii} emission
from all  three ionization ranges listed above.  The most unusual Sy2 sources
are IRAS 18325-5926 and MCG-5-23-16 where very broad lines were detected.  The
former source shows a broad (spanning 4.5--7 keV) profile,  red (low energy)
asymmetry and narrow peak on the blue (high energy) side centered at 6.8 keV. 
It is unclear whether this is a single line or a blend of several features but
it is reasonably certain that the extended emission redward of 6.4keV is real.
MCG-5-23-16 shows a broad component extending from 5--8 keV with a narrow peak
near 6.4 keV. Multiple components are again a distinct possibility. The
``hidden'' quasar IRAS 09104+4109 (Fabian et al.\ 1994b) has been proposed as
an ultraluminous example of the Sy2 class. It shows a very strong but narrow
{\sc Fe} line centered at 6.65 keV.

\subsection{BLR H$\alpha$ Statistics and Comparison with \feka\ \label{stat}}

We have obtained, or gained access to, optical spectra for almost every 
Seyfert 1 galaxy that shows a strong \feka\ (ASCA) detection so far. 
Observations that have not been presented elsewhere are summarized  in Table
2. The format is as follows: Column 1 - abbreviated IAU designation; Column 2
- other name;  Column 3 - date of observation; Column 4 - universal  time of
observation; Column 5 -  exposure time  in seconds; Column 6 - observatory;
Column 7 -  telescope; Column 8 -  spectrograph; Column 9 -  grating and
Column 10 - resolution FWHM (in \AA).

Table 3 summarizes the \hiha \  profile properties for both old and new 
observations. The format is as follows: Column 1 - IAU code name; Column 2 - 
source name; Column 3 - Seyfert type; Column 4 - EW in \AA; Column 5 - FWHM in
km/s;   Column 6 - uncertainty at  2$\times \sigma$\ confidence level;  Column
7 -  centroid line shift (at half maximum) in  \kms; Column 8 -  line centroid
uncertainty; Column 9 - reference code.  Uncertainty of H$\alpha$ EW values
should be $\approx$ 10\%.  Figure 2 shows a comparison of  H$\alpha$ and
\feka\ profile  FWHM values. The values plotted for  \feka\  come from the
(2.35$\times$) $\sigma$ values given in Table 1. They were derived from the
broad line  fits to the spectra (see Table 1 references). They were usually
estimated  by fitting the X-ray continuum with (a) a nonthermal  power-law,
(b) a  soft X-ray absorption law and (c) a hard reflection component. In
general  inclusion of component (c) enhances the continuum underlying the
line  resulting in a reduction of the derived EW and FWHM values. In some
cases   plausible solutions can cause the line to become unresolved  (e.g.\ 
MCG-02-58-22).

Figure 2 shows  that \feka\ line is systematically broader than \ha\ in all
{\em radio-quiet}  AGN. In  the case of MCG-6-30-15  (dots in Figure 6) and
NGC 4151, FWHM(\Ka)/FWHM(\ha) $\approx$ 40 and 10 respectively.  The  limited
data for radio-loud sources shows a smaller difference in FWHM. 
For  the  Seyfert 1 galaxies listed 
in Table 3 we find:  (1) the hypothesis that \feka\ and  \ha\  FWHM come 
from the same distribution is rejected at a  confidence level $\simgt$ 0.995 
using generalized (including upper limits for \feka) Wilcoxon tests and  
(2) \feka\ and \ha\ FWHM are not significantly correlated. A Kendall test  
with the inclusion of censored data yields a probability P$\approx$ 60\% 
that a correlation is absent. EW values of \feka\ and \hiha\ are also not 
significantly correlated. The Spearman correlation coefficient is 
$\rm r_S \approx 0.31$. For 25 data points the probability to have this 
value in a random sample of observations taken from an uncorrelated parent 
populations is 12 \%. Restricting  attention to the objects studied by  
Nandra et al.\ (1997), we do not find any significant correlation 
for half  maximum centroid shifts.  The most interesting trend emerging 
from a lineshift comparison is that the objects  with a strongly redshifted 
broad component in \feka\ show symmetric (unshifted) and relatively narrow  
\hiha\ line profiles.  

\subsection{Possible Sources of \feka\ Line Emission}

Our considerations of origins  for \feka\  are focused on the Seyfert galaxies
where the bulk of  the \feka\ spectral detections of reasonable S/N are found.
In a unification driven approach  we assume that:  (1) a  supermassive black
hole and associated accretion disk lie at the center of each AGN, (2) Seyfert 2
galaxies  (at least some of them)  are  Seyfert 1's seen edge-on
(i$\simgt$45$^\circ$) and (3) the optical broad line emission region in Seyfert
2's is  obscured by the putative cold dust torus. In this framework we can 
identify at least six possible origins for \feka\  line emission  (listed
roughly in order of decreasing distance from the black hole): (A)  cold \feka\
emission from the obscuring dust torus which is  assumed to be optically thin
at $\sim$ 6keV in most cases (see e.g.\   Mushotzky et al.\ 1993: Ghisellini,
Haardt \&\  Matt 1994; Weaver et al. 1996); (B) \feka\ emission from the
classical BLR; (C)  \feka\ emission from a photoionized scattering  region (a
warm absorber); (D) cold \feka\ emission due to fluorescence reflection (or
emission) from the accretion disk; (E)  \feka\  emission from  highly ionized
 coronal gas. Coronal gas (T $\simgt 10^9~$K) can give rise to strong \feka\
 emission at 6.67 or 6.97 KeV; (F) cold
\feka\  emission from a (non-disk) cloud distribution  (not necessarily
responsible for optical/UV LIL and/or HIL). Option (A), (B), (C), and (E) are
briefly discussed below, as they appear rather unlikely to be the dominant
source of  \feka\ emission.  Option (D) and (F) deserve a much deeper
discussion, deferred to \S\  \ref{disk} and \S\ 4 respectively. Our model
considerations  focus on MCG-6-30-15 as the best studied source with  a
characteristic broad \feka\ line profile.  The broadband spectrum of
MGC-6-30-15 is consistent with a typical AGN continuum as described  by  Mathews \&\
Ferland (1987; MF87) except that the sub-millimeter break may occur at higher energy
than  implied by their continuum ($\simlt$1$\times10^{-3}$ Ryd in MCG-6-30-15).
We nonetheless assume that the sub-millimeter break occurrs at
1$\times10^{-5}$ Ryd  as parameterized by MF87.  The ionizing luminosity 
of MCG -6-30-15 is L$_{\rm ion} \approx 6\times 10^{43}$ ergs
s$^{-1}$ from $\rm \log \nu
L(\nu) \approx 29.0 $ ergs\ s$^{-1}$  at $\nu  \approx 14$\ Hz
($\rm H_0 = 75 km ~~s^{-1}
~Mpc^{-1}$, and q$_0 = 0$\ is assumed throught the paper; the corrected heliocentric radial
velocity of MCG-06-30-15 measured on our spectrum is $\rm v_{r,h} \approx 2317$ \kms,
which implies a distance $\approx$ 30.9 Mpc.)

 Table 4 reports expected line luminosities of MGC -6-30-15
for cases A, B. C, E. In Column (2) we report the \feka\ luminosity computed in the case of 
gas with
solar chemical abundances, while in Column (3) we report the \feka\ 
luminosity  in the case of a factor 10 iron overabundance with respect to the solar value. 
In Column (4) the dominant
ionization stage of iron is reported. Column (5) lists the expected \hiha\
luminosity. Luminosities are in ergs s$^{-1}$. These values should be 
compared to an observed total \feka\ luminosity $\rm \log L($\feka$) \approx$ 41.2,
and to $\rm \log L($\hiha$) \approx$ 40.8\ .  The L(\ha) value has been corrected for galactic
extinction assuming $\rm A_B \approx 0.15$, but not for internal absorption. 
The \feka\ narrow component luminosity 
is $\approx$ 0.3--0.5 of the total L(\feka). 

\subsubsection{Emission from a Dusty Torus}

Option A would only give rise to a narrow \feka\ line centered near rest 
energy. Strong variations observed in the ``narrow'' 6.4 keV part
of the MCG-6-30-15 spectrum complicate this interpretation (Iwasawa et al.\ 
1996b). For MCG-6-30-15 there is also a photon deficit problem. Table 4 reports
the results of explorative {\tt CLOUDY} (Ferland 1996) photoionization computations. A simple, optically
thin model of a dusty torus, located between 3 and 10 pc, and photoionized by
the central continuum source, produces a \feka\ luminosity that is more than a 
factor $\sim 100$\ below the observed \feka\ luminosity. This remains true even if the attempt to
ascribe to torus emission the ``narrow component'' only of the \feka\ line.
The situation may be markedly different in Sy 2 galaxies, where a narrow \feka\
line is observed against a strongly absorbed X-ray continuum (Weaver et al.
1996; Ghisellini, Haardt, \&\ Matt, 1994; Krolik, Madau \&\ \.{Z}ycki 1994).

\subsubsection{Warm Absorber}

The warm absorber  (option B) is also unlikely to be the dominant source of
\feka\ line emission  (Reynolds \&\ Fabian 1995; see Table 4).  The
photoionization calculations presented here are based on the model by Reynolds \&\ Fabian:
gas with T$\approx 10^5$K, $\rm n_e \sim 10^6$ \cm3, and moderate column
density $\rm N_c \sim 10^{22} cm^{-2}$, located within the BLR or at the outer edge of
the BLR is photoionized by the AGN continuum. This model is not the only one
proposed for the warm absorber (e. g., Mathur, et al. 1994), but it is probably
adequate for low luminosity radio-quiet AGN.   There are two
main obstacles to ``warm iron'' \feka\ emission. (a) a photon deficit; (b)  the
effects of Compton scattering on the line profile (Fabian et al.\ 1995).  As a
central corona is likely to be Compton thick, extreme broadening may occur
($\sim 1$ keV; Fabian et al. 1995; Nandra et al.\ 1996). 

\subsubsection{Classical BLR}

The main problem with \feka\ BLR emission are that (1)  there is a flux
deficit,  unless high iron abundances are invoked (see Table 4),  and (2) \feka\ and Balmer
line width are inconsistent at least in Seyfert 1 galaxies (\S\ \ref{stat}). 
Both problems are solvable if only the narrow component of the \feka\ line is
ascribed to the BLR. Our photoionization simulations show that, if iron
abundance is 10 times the solar value, then the luminosity of \feka\ produced
within the BLR  agrees with the observed luminosity of the \feka\ narrow
component. 

The so called ``X-ray Baldwin effect'' (anticorrelation between \feka\ EW and
X-ray luminosity) has been interpreted as  evidence for a common origin of
\feka\ and \civ\ emission in the BLR (Iwasawa \&\ Taniguchi 1993). The ASCA
data do not support such an anticorrelation. The EW  of \Ka\ is strongly
dependent on the fit of the X-ray continuum. If we consider the Iwasawa \&\
Taniguchi (1993) objects for which  ASCA observations are available (16 AGN,
mainly low-luminosity), and that were fitted with a reflection component
whenever appropriate, we find that the anticorrelation disappears  (a Kendall's
$\tau$, generalized for the inclusion of censored data, yields $\approx$ 15 \%\
probability of no correlation).  We must point out, however, that if we 
substitute the ASCA data, and keep the other  Iwasawa \&\ Taniguchi (1993)
values (a total of 38 objects), obtained from GINGA observations, then the
correlation does not go away (Kendall's $\tau$  $\Rightarrow$ P $\approx$
0.003--0.007). As the resolution of the GINGA data was substantially lower, and
the typical profile in Sy 1 galaxies resembles the one observed in MCG -06-30-15,
with a broad redshifted component and a narrower component at 6.4 keV,
a significant part of the broad companent may have been lost in the fit of
the underlying continuum. The apparent lack of any 
``X-Ray Baldwin effect'' in the ASCA data is therefore consistent with the idea  
that the \feka\ narrow component only is emitted within
the BLR (Marziani, et al. in preparation). 

\subsubsection{Hot Corona}

A contribution from ``hot iron'' (more than 17 times ionized)  cannot  be ruled out for Seyfert nuclei
because a high energy residual is observed for most \feka\ disk candidates
(see  discussion in Section 4).  We will show that, in fact,  hot  iron emission is
expected from a photoionized disk. However, the most robust result of ASCA
observations is that hot gas cannot be the dominant source of \feka\ emission
in Sy 1 galaxies, as the line energy is almost always around 6.4 keV,
implying low ionization for iron (Nandra et al. 1997a).  Nevertheless, coronal gas at a temperature close to the virial temperature T$_{\rm vir}
\approx 5\times 10^{11}$ M$_{\rm 8}$ r$_{14}^{-1}$ K, within r $\simlt$
10 --100\rg\ (where the gravitational radius \rg\ = GM/c$^2$) 
should be a strong \feka\ emitter. The resulting \feka\ luminosity
is strongly dependent on the amount of matter that makes up such a corona. In
Table 4, we compute a model assuming T=10$^9$K, and $\rm N_c \approx 10^{25}
cm^{-2}$. Redshifted hot coronal gas in addition to BLR emission provides an
alternative to accretion disk emission. The observed profile may be reproduced
equally well.  The combined effects of  gravitational redshift and
increasing rest energy can produce centroid  values $\simgt 5.0$ at 6 \rg\
(6.19 keV at 20 \rg) if the ionization stage of  iron is  high. However, this 
requires some sort of fine tuning.  

\section{Broad \feka\ and HI H$\alpha$ Accretion Disk Models \label{disk}}

Two independent lines of recent work consider emission from an accretion disk
for \feka\ (\.{Z}ycki \&\ Czerny 1994; Matt, Perola \&\ Stella 1992; Fabian et
al. 1995; Dabrowski et al. 1997) and for \hiha\ (Dumont \&\ Collin-Souffrin 1990a,b \&\ Collin-Souffrin \&\ Dumont 1990;
Rokaki, Collin-Souffrin \&\ Magnan 1993, Eracleous \&\ Halpern 1994). The first
order fits to the \feka\ line are qualitatively more  satisfactory than fits to
optical profiles. Model profiles computed for a rotating disk surrounding a  Schwarzschild or 
Kerr black hole are most clearly consistent with the \feka\ line profile 
observed in MCG-6-30-15 (dashed profile in Figure 6 shows Tanaka et al (1995) 
fit).  However, as the number of \feka\ detections increase, and
line variability measures become more accurate, the disquieting trend is that
model fits become poorer and require an ever larger  range in the parameter
space for disk models (e.g.\ Iwasawa et al.\ 1996b; Weaver et al.\  1997).
There is no evidence for  convergence toward some characteristic disk model. 
Only a few other sources (Fairall 9, NGC4151, NGC5548, IC4329A) show structure unambiguously similar  to
MCG-6-30-15.  This may be due to real profile differences  or to the large
variations in S/N of the data. The latter possibility is likely because the
composite profile constructed by Nandra et al.\ (1997) from about 16 objects 
(see solid curve in Figure 6) is qualitatively similar to the MCG-6-30-15 
profile.  To summarize, one can fit almost any \feka\ line profile with  a  disk model
especially: (a) when one fits low S/N  X-ray line profiles to models where most
of the emission is produced  at the innermost disk radii and (b) when the
entire parameter space of Schwarzschild and Kerr models is available.  Coupling
this with assumptions about an emissivity law allows one to fit a disk model to
any profile.  This implies that a greater degree of self consistency is needed. 

In the optical, where the lines are better defined the problems are basically the same, with
the additional difficulty of {\em very} poor agreement between models and observations. 
Objects like Arp102B are rare--forcing  disk emission advocates to argue that (1) objects
with similar double-peaked profiles are most often obscured, and/or  (2) that most emitters
produce the line at larger radii in the disk (making the discovery of Arp102B somewhat
miraculous).  Truncation of line emission at r $\approx$ 1000 \rg\ demands a physical
motivation that has yet to be found (see Eracleous \&\  Halpern 1994). 

Nonetheless, an accretion disk origin for the {\sc Fe} line emission might
offer a  straightforward explanation for some important observational 
findings.  If  \feka\ is emitted extremely close to the central black hole
then that line  emission is most sensitive to  disk  structure. This might
explain why a  number of intermediate  redshift (z $\sim$ 1) quasars show
neither  strong \feka\ nor a strong   reflection continuum component (Nandra
et al. 1997c). Disk structure could be markedly different in  radio loud AGN
as well. The standard \a --disk theory suggests  that the region dominated by radiation
pressure may not exist if  the dimensionless accretion rate \.m $\simlt$ 0.4
(Shakura \&\ Sunyaev 1973; Frank, King, \&\ Raine 1992). This  case corresponds to the lowest luminosity
AGN, which are thought to radiate in the sub-Eddington regime. In  
intermediate redshift quasars, the dimensionless accretion rate may approach
unity, invalidating the thin disk  assumption, and the central regions of the
disk may puff up (Abramowicz, Calvani, \&\ Nobili 1980; see also Calvani, 
Marziani, \&\ Padovani 1989 for a review). In this case, we expect that \feka\
is emitted only in the narrow funnel of the inflated disk, as its external layers would become too cool (Madau 1988).  This would explain the rare
detections of \feka\  in intermediate and high redshift quasars, including
both the  radio-loud and radio-quiet variety, and observed dependence on
luminosity of the line profile (Nandra et al.\  1996; Nandra et al. 1997c).
If  \feka\ is emitted in the funnel, then it can only be seen if the disk is
observed nearly face-on, the rarest of the  occurrences.  It is intriguing
that the only quasar where strong \feka\ is detected, PG 1116+215 (Nandra et
al.\ 1996),  fits very well with this scenario. Model fitting of \feka\ suggests i
$\approx$ 0$^\circ$.  The very narrow \hiha\ profile, and  the extremely
strong \feii\ emission make it a close replica of I Zw 1. These features can
be explained in terms of emission  from a face-on disk (Marziani et al.\
1996).

LIL, HIL and \feka\ might all arise from the  same  component with X-ray, UV
and optical line emissivities peaking at different  radii from the   central
continuum source.     The FWHM difference  described in \S\ 2.3 could be 
accounted for if  the X-ray lines were produced in a disk with \feka\ produced
near the center  and the Balmer lines further out. At first glance the lack of
a correlation  between \feka\ and \ha\ FWHM appears to pose a problem for 
models of this  kind.  This is not necessarily the case however because in
disk models the  width of \ha\ (FWHM) is  driven by illuminating flux while
FWHM \feka\ is  sensitively driven by the inner disk emitting radius.  

%
%
%
%
%
%

\subsection{A Disk Illumination Model for Both H$\alpha$ and \feka}

In MCG -06-30-15, the Eddington ratio is $\eta_{\rm
Edd}  = {\rm L}_{\rm bol}/{\rm L}_{\rm Edd} \simlt$ 0.34 ($\dot{\rm M} \approx
2\times 10^{23}$ g s$^{-1}$),  consistent with the assumption of a thin disk
down to the innermost stable orbit. We estimate a lower limit to the mass of the central black hole 
M$_{\rm BH} \simgt 4\times 10^6 M_\odot$\ from the assumption of cloud virial
motion, following Padovani (1989).  We assume  M=$10^7$\msol\  and $\dot{\rm
M} \approx 2\times10^{23}$ g s$^{-1}$\  in the computation of the disk
structure (if line emission takes place in  a disk the mass estimate should be
divided by $\sin(<\rm i>) \approx 0.5$).  We assume that this is appropriate
for MCG-6-30-15 and all low-luminosity Seyfert 1's studied in this work.

We model the region where the continuum  is produced as a  sphere of radius
r~$\approx$ 100 \rg\ with $\tau_{\rm es} \approx$ 0.25. Radiation emitted in
this region directly illuminates the disk, and  is also partly scattered 
towards the disk by the free electrons in the sphere. In  addition, a
spherical  halo of free electrons may surround the disk, up to 10$^6$\ \rg, 
again with $\tau_{\rm es} \approx$ 0.25. Given the large uncertainty in
the ionizing luminosity of AGN, we compute  models for two extreme choices
of the continuum: (a) the observed continuum, according to MF87. 
This choice probably overestimates the ionizing luminosity, as it
includes the big blue bump and the reflection component that are thought to be
produced within the disk; (b) a broken power law with slope $-1.4$ between
1$\times10^{15}$ Hz and 50 keV, and $-2.5$ for energy $\simgt$ 50 keV, which may
yield an underestimate of the ionizing photon flux. 
Computation of the scattered flux as a function of disk
radius is as carried out by Rokaki et al. (1993). Relativistic effects (most notably
returning radiation i. e., radiation emitted by an accretion disk which returns
to its surface due to gravitational focusing or the shape of the disk;
Cunningham 1976) are neglected  in the scattering process, even if they are
thought to be relevant in the innermost region of the disk.  Returning
radiation may enhance \feka\ emission by a factor $\approx$ 2 (Martocchia \&\
Matt 1996; see also Dabrowski et al. 1997). 

We consider here the radial disk structure appropriate for a Kerr metric, with 
the angular momentum per unit mass a = 0.998 M  (c = G = 1; as shown by Thorne 1974,  
the radiation emitted by the disk and swallowed by the hole prevents spin up beyond a limiting
state a/M$\rm _{lim} \approx 0.998$ M; the corresponding innermost  stable orbit is at 
 r$\approx$ 1.25 \rg) and use the 
equations for the radial and vertical structure of a  standard \a -disk 
model, with the corrective terms  given by Novikov and Thorne (1973).  The
disk  is  divided  into concentric annuli of fixed  fractional width $\Delta
{\rm r}/{\rm r}\approx 0.2$. Surface density, column density,  and 
illuminating flux are assumed constant within each  annulus.  We approximate the hydrostatic equilibrium (leaving the vertical
structure unperturbed by the incoming radiation) assuming that the  density
within the disk increases linearly with column density. This is a crude
approximation since the vertical structure of the disk should be recomputed to
take into account the heating provided by the illuminating flux.
In the outer layers of 
the innermost regions, if the density is too low and the radiation density
too high (for values of $\rm \xi = L_{ion}/r^2 n_e \sim 10^4$ ergs s$^{-1}
$cm),   the gas is heated to T $\sim 1\times10^{10}~~$K. In this case
the disk layer actually becomes part of a corona, and   we assume that
radiation penetrates to a layer within the disk of optical depth larger by 
$\Delta \tau_{\rm disk}  = 1$, 
with  flux reduced by a factor $\exp (-\Delta \tau_{\rm disk})$.   We employed
the  photoionization code {\tt CLOUDY} (Ferland 1996) to compute the
emissivity in  the \feka\ and Balmer lines. 

\subsection{Illumination Model Results}

\subsubsection{\feka \label{emission}} 

We identify four vertical zones within the disk. (a) Very hot gas -- a  halo
that has undergone a thermal runaway due to heating by scattered   radiation in
the lowest density layers of the disk ($\tau < 1$). (b) Warm gas in an ``intermediate'' region
of higher density, whose  electron temperature varies from  $\simgt$10$^7$ down
to a few $10^4~~$K.  (c) Ionized gas at a few $10^4~~$K; (d) cold,
almost neutral gas  which is less heated by the incoming radiation (T $\rm
\simlt few \times ~10^3~~$K).  Reflection models usually suggest that the
``cold" \feka\ emission  is  produced in the
two  deepest layer. In Table 5 we report the results of our calculations. In
Column (1) we give an identification number for each model. In Columns (2) (3),
(4), (5) the inner and outer radii, and the electron scattering optical depth  for
the inner sphere and the outer halo are reported. In Column (6) we list the shape of
the continuum (as described in the previous section). Column (7), (8), and (9)
yield the logarithms of the \feka\ luminosity from cold iron emission, 
the ratio L(\feka)$\rm_{Hot}$/L(\feka)$\rm _{Cold}$, and  the $\rm \log(L($\hiha))
respectively.  

Our model of the vertical disk structure  allows us only a coarse estimate of the relative
flux contributions from ``cold"  and ``hot" iron. The computed ratio  {\sc Fe}$_{\rm
Hot}$/{\sc Fe}$_{\rm Cold}$ ($\approx 0.1-0.3$)  is   sensitive to returning radiation which 
is not taken into account. The predictions of our model  for  {\sc Fe}$_{\rm Hot}$/{\sc
Fe}$_{\rm Cold}$  are nevertheless in qualitative agreements with those of  Matt, Fabian \&\
Ross (1996).   We are also aware that mean escape
probability formalisms are not adequate for the treatment of Compton scattering and radiative
transfer of the \feka\ line. Nevertheless,  our
photoionization calculations for cold \feka\ agree  with the results  of Monte Carlo
simulations of \feka\ emission from a homogeneous layer of  cold matter by George \&\ Fabian
(1991). They find that  $\approx$ 6\%\ of the photons between 9 and 30 keV are converted
into  \feka\ photons. This yields L(\Ka) $\approx 1.2 \times 10^{41}$ ergs 
 s$^{-1}$ for 
MCG-6-30-15 (the observed total \feka\ luminosity is 
L(\feka) $\approx 1.6 \times 10^{41}$ ergs s$^{-1}$).

From Table 5 we deduce also that the observed luminosity values for MCG
-6-30-15 are in the range predicted by our disk models. We must remark that
the  choices of the continuum are probably extreme (new observations suggest
that the ``big blue bump'' may not be as prominent as previously thought; Laor
et al. 1997).  The MF87 continuum gives rise to line luminosities somewhat
larger than observed, as the observed L(\Ka)  of MCG -6-30-15 in the ``medium
phase'' is $\approx$1.6$\times  10^{41}$ ergs s$^{-1}$ (Isawasa et al., 1996). 
On the other hand, the ``power-law'' continuum has an ionizing luminosity a
factor $\sim 10$\ lower than that of the MF continuum, for the same optical
flux, and it therefore not surprising that fluxes are lower than observed. The
total \ha\ luminosity is rather sensitive to the presence of an outer halo, 
but the \feka\ luminosity is not, as emission in the innermost disk region
$\simlt$ 20 \rg\ is by far the strongest. Figure 3 shows emissivity profiles
for \hiha\  and \feka\ derived from our models.  

Next we computed the cold \feka\ line profile  following Fanton et al.\ (1997),
and using the emissivity law obtained from Model 1. The code developed by  Fanton et al.
computes the trajectories in the full Kerr metric taking  into account all the relevant
general relativistic effects.   The computed profile is therefore accurate at any distance
beyond the event horizon and for disks seen at any inclinations. Integration was carried out
in between  1.25 \rg\ $\simlt$ r $\simlt 20$ \rg. Figure 4 shows the results of our  fits to
the MCG-6-30-15 data (Tanaka et al.\ 1995) where we were able to satisfactorily reproduce
the  \feka\ profile.

\subsubsection{\hiha} 

The vertical stratification of density implies a variety of  physical
conditions within the disk. This has the important implication that the disk 
is able to produce Balmer line emission at almost any distance from the 
continuum source, in regions (c) and (d) of \S \ref{emission}. The \ha\
emissivity law for Model 1 can be approximated with a  shallow power law (p =
$-0.71$)  for $10^2$ \rg\ $\simlt$ r $\simlt 10^4$ \rg\ and with a steeper law 
(p=$-2.33$)  for r $\simgt 10^6$ \rg. This emissivity law makes the production
of symmetric  single-peaked profiles very likely, as the emissivity  favors
emission from  the outer regions of the disk, where relativistic effects are
negligible.  $\rm \epsilon(r)\times r^2$\ peaks at r $\sim 10^4$ \rg,  a value
that is in agreement with the average  distance for  Balmer line emission
deduced from  reverberation mapping. Fig.\ 5 shows superimposed \ha\ and  disk
profiles based on model 1 for 15 Seyfert 1's  included in the  Nandra et al.\
(1997a) sample. Model  profiles were computed following Chen \&\ Halpern (1989)
rather than Fanton et al. (with the  emissivity law of Model 1) as their weak
field  approximation is adequate for the relativistic treatment at large
radii.  Best fits were achieved using the downhill simplex algorithm by Nelder 
\&\ Mead (1978). The outer radius of the emitting region was allowed to vary
within a narrow  range, $10^5$ \rg $\simlt \rm r_{out} \simlt$  $10^6$ \rg\ 
(the distance at which disk fragmentation due to self gravity is expected to
become relevant). An additional free parameter is a scale factor. In Table 6 we
report the results of \hiha\ disk model fitting. The format is: Column 1 and 2
- IAU code and common name; Column 3 -  $\rm r_{out}$; Column 4 - \hiha\ disk
inclination from our model 1 fits;  Column 5 - the reduced $\chi^2$; 
Column 8 - Column 6 and 7 inclination and reduced $\chi^2$\ for Model 2 fits
(not shown in Fig. 5); \feka\  disk inclination from Kerr model fits  (Nandra et al.\ 1997). 
H$\alpha$ model fits of 3C 120. Statistically good fits
are the ones with $\chi^2 \simlt 1$.  The formal uncertainty in the estimate of the
inclination is typically a few degrees (it is undefined if the fit is not
statistically good), but the actual uncertainty is probably much larger.  
The distribution of scattering matter affects the estimate of the inclination,
since it change the turning point in the slope of the emissivity laws (Fig. 4).
Model 1 and Model 2 should  be regarded as limiting cases for the emissivity
law.

\subsubsection{Main Results and Implications}

Our attempt to constrain the disk models has met with mixed success. 
We summarize here the main achievements and problems. 

\begin{itemize}

\item There is poor agreement between the disk inclination derived from X-ray
and optical lines for MCG-6-30-15. This is an especially disturbing case. The
extremely narrow \hiha\  profile requires small inclination while the broad
\feka\ profile  constrains  it to $\approx 30^\circ$, even in the case of a
Kerr black hole with a$\approx$M. Disk inclination is the only free parameter
for our model because the inner  and outer emitting radii are constrained by
the emissivity law.  $\sim$ 7 objects studied by Nandra et
al.\ (1997a,b) have  derived \Ka\ disk model inclinations well enough
constrained for  comparison. The small number precludes a  reliable statistical
analysis. Even including some objects with uncontrained inclination from \feka\
data,
we find a statistically good fit for \hiha\ and 
inclination consistency between \hiha\ and \feka\ only for NGC 3227, NGC 3783, Mark 841,
NGC 4593, NGC 7469 and NGC 5548.

\item Disk models do not reproduce most Balmer line  profiles of Seyfert
galaxies. This is most likely true for all models computed in this paper. There
is no known physical reason to limit the  emitting radius of \hiha\ except for
the outer radius at which  the disk fragments ($10^5$ - $10^6$ \rg).
Integrating within this limit,  line cores are well reproduced only when they
are symmetric and  unshifted. The shape and asymmetry of line wings also does
not agree  with disk model predictions (Romano et al.\ 1996).

\item The comparison between observed and disk model \hiha\ profiles suggests 
a simple variant of the old cloud model  (see e.g.\  Mathews \&\ Capriotti 
1985, and references therein)  capable of explaining the new data: a flattened 
ensemble of dense ($\rm n_e \sim 10^{10}$ to $\sim 10^{13}$\cm3)  clouds,
co-axial with the accretion disk,  spiralling inward (or outward) because of
drag forces, on a time-scale much shorter than the radial drift time scale for
an $\alpha$-disk. Cloud production might occur in a wind  from the disk
surface, or the wind itself could be a source of \hiha\ and \civ\ line emission
(e. g., Murray \&\ Chiang 1997). 

The real challenge for any BLR model is not the ability to reproduce one single
profile, but rather to account for the distribution of line width, shift and
asymmetry observed in any sample of AGN. No model proposed until now has
achieved this goal.  Asymmetries may be  produced because the clouds
are optically thick to the lines they emit (i.e.\ the lines we observe come
mostly from the illuminated face of the cloud), and because of the spiralling
path. If the system of clouds is observed nearly face on, then we have
symmetric, unshifted, profiles, and the narrowest of all observed profiles.
Larger asymmetries are expected  if the line is broader (as observed): in this
case line emitting clouds have a larger component of their non-circular
velocity along the line of sight. 

\item The {\sc Fe} profile shape may be complicated by a significant 
and variable contribution from hot {\sc Fe}. This may help account for 
the large observed EW's and provides an additional free parameter for 
adjustment to the profile shape. In particular it might account for 
the  wing on the high energy side of the line observed in sources like MCG-5-23-16 
(Weaver et al.\ 1997). 

\end{itemize}
                                                                                                                                    
\section{Other Models for \feka\ Emission}

\subsection{Major Challenges for Disk Models}

An important uncertainty, that affects our interpretation of the {\sc Fe} line
origin, is whether it arises from a single emitting region. Almost all past 
disk model fits, including the previous section,  have 
treated the \feka\  line as a single component. There are several forms of
evidence to support the hypothesis that the line is composite.

1) Many Seyfert 1 spectra show a narrow \feka\ peak very close to the rest
energy of 6.4 keV. This is  not a problem for a wide range of disk models and
is, in fact, consistent with the  expectation of a Doppler-boosted blue peak
from an illuminated disk viewed at intermediate  inclination.  The fact that
the blue peak always occurs at 6.4 keV would  just be a coincidence. However
the narrow feature can also be  reasonably ascribed to an unshifted
independent component. This possibility is strengthened because  some of the
best line profiles, including MCG-6-30-15, show a smooth high  energy (blue)
wing. Nandra et al.\ (1997) present composite spectra  (see solid curve in
Figure 6) made both  with and without the two ``best'' examples, MCG-6-30-15 
and NGC 4151. Both  composite spectra show a similar blue wing on \feka\
suggesting that  it is a common property of Seyfert 1 line profiles. The range
of disk  models explored in the literature almost always show a sharp drop on
the blue side of the profile which is inconsistent with this blue shoulder.
The recent detection of MCG-5-23-16 (Weaver et al.\ 1997 -  not included in
the Nandra et al. 1997 study),  exacerbates this problem showing perhaps the
strongest blue shoulder yet observed. An obvious decomposition of the line
into two symmetric Gaussian components is: (1) a narrow component centered at
$\rm E_n$=6.4$\pm$0.05 keV with FWHM=0.25 keV and (2) a broad component
centered at $\rm E_b$=5.9$\pm$0.1 keV with FWHM=1.6 keV. The blue wing appears
to be a natural extension of the second component under the first rather than
an independent feature. 

2) Recent variability results for the best studied Seyfert 1 MCG-6-30-15 show
that the narrow ``blue'' (6.4keV) and broad red components vary independently
(Iwasawa et al.\  1996b). The blue peak responds quickly to  continuum changes
while the red peak responds more slowly or remains relatively constant. This is
again consistent with the notion  that we are observing two independent
features. The variability studies  add another problem for disk models. 
Pre-ASCA variability studies (Mushotzky et al. 1993) already found potential
problems with disk models because the \feka\ line did not show the relatively
short time scale response to continuum changes that is expected from simple
disk models. This issue is discussed in a companion paper (Sulentic et al.
1997).


3) Many objects show breaks in the red wing that could indicate discrete
components.  Only higher s/n observations will decide if  the red wing is
continuous and consistent, in detail, with disk models.  The recent \feka\
detection in MCG-5-23-16 shows clear evidence for two or  three components (one
at 6.4 keV as discussed above) (Weaver et al.\ 1997). Model fits to the broad
feature in MCG-5-23-16 leave a significant residual between 5-6 keV and
possibly also on the blue side. A 5.5 keV component was also discussed for NGC
4051 (Guainazzi et al.\  1996) and 3C 109 (Allen et al.\ 1997) however the
latter reference suggests  that this may be related to an instrumental
artifact. 

4) Finally there is the problem raised by unification models that view (at
least some) Seyfert 2 galaxies as edge-on Seyfert 1's where the BLR is obscured
by a dust torus (see also Turner et al. 1997 and Nandra et al. 1997). Seyfert 2
galaxies are most probably an heterogeneous class, but there is evidence that
Seyfert 2 with \feka\ detections may be genuinely obscured Seyfert 1 (see e.
g.,  Dultzin Hacyan 1997, and private communication). 
Emission from an inclined  disk would produce  broad
profile which are easily distinguishable.  Only Seyfert 2 sources IRAS
18325-5926 and MCG-5-23-16 can be reasonably well fit with disk models. The red
asymmetry and blue-shifted blue peak are consistent with a reasonably high
inclination (i=40-50$^\circ$: Iwasawa et al.\ 1996a) as expected for a Seyfert
2.  The current status of disk emission models for \feka\ lines in Seyfert 2
objects hearkens back to the situation for a disk signature in optical broad
lines as discussed in the introduction.  If we see an obvious disk signature in
IRAS18325-5926 (and MCG-5-23-16) then where is the corresponding signature in
other Seyfert 2 objects?  Why we do not see an edge-on disk profile in NGC
1068 (this object may be optically thick out to 20 keV: Mushotzky et al.\ 1993),
NGC 3147 and 6552 as well as Markarian 3 and other Seyferts 2's 
observed by ASCA? 
 The lack of coherence in the model fits to different AGNs of the same
subclass is already disturbing.  The large equivalent width of the
IRAS18325-5926 detection (and now also MCG-5-23-16), if from a disk, suggests
that many other inclined (but not edge-on) disks should be detectable.


\subsection{Alternatives to \feka\ Disk Emission \label{alt}}

Alternative models for the complex emission profile near 6.4 keV (or just the
broad component) must avoid the difficulties that were summarized in the
preceding section. They must also satisfy the following general requirements: 
(a) produce strong \feka\ emission (EW$\sim$100-200eV), (b) broaden it
($\sigma\sim$0.6keV), (c) red shift the bulk of the emission and (d) to a
lesser extent, symmetrize the line (assuming that the narrow and broad
components are independent). It must also account for the FWHM difference
between \feka\ and the UV/optical HIL/LIL if a BLR unification is attempted.
The first task can be accomplished with an ensemble of emitting clouds
provided that the column density is high enough and the covering factor is in
the range 55-85\% (Sivron and Tsuruta 1993; Nandra \&\ George 1994). The
strength of the line is a problem for disk and non disk models alike. A
photoionized  medium  is one of the solutions previously proposed for this
problem (e.g.\   Tanaka et al.\ 1995). The line profile can be broadened most
easily by increasing the velocity dispersion of the cloud ensemble but
possible constraints imposed by the sharpness of absorption edges (Fabian et
al.\  1995) motivate us to look for solutions that minimize Doppler
broadening.  Several other possibilities have been considered, some of which
could also introduce the required red displacement that is observed (e.g.\ 
Fabian et al.\ 1995) by (1) Comptonization, (2) transverse Doppler effect and
(3) gravitational redshift.

The difference in average profile FWHM between \feka\ and \hiha\  greatly
constrains non-disk models where \feka\ arises in the optical broad line
clouds.  Our previous work on modeling optical line profiles with bi-conical
emission structures (as alternatives to disk models; Zheng, Binette, \&\
Sulentic 1990) leads us to consider that possibility also for {\sc Fe}
emission.  A bi-cone model might provide an explanation for the strong changes
observed in the narrow component of MCG-6-30-15. As the approaching side of
the bi-cone should be the one closest to the observer, one could see
variations following continuum variability with essentially zero time delay. 
This is especially true if the opening angle of the cone is small. Dense
clouds (\ne $\simgt$ 10$^{13}$\cm3 are a likely producer of \feka, and a
continuum reprocessor as well (Guilbert \&\ Rees 1988; Lightman \&\ White 1988) giving a flattening of
the X-ray spectrum similar to the situation for a disk reflection component. 
Such dense clouds should not be a strong source of Balmer line emission.

%
%

\subsubsection{A First-Order Model}

We made a first-order test of the possibility that the \feka\ line 
emission is produced in a cloud ensemble surrounding the central 
source. Earlier work on a cloud model was discussed in the previous 
section. Our extension of this work is intended to show only that a 
biconical cloud distribution can reproduce the most complex \feka\ line 
profiles. Options include  modeling: (1) the  entire {\sc Fe} line 
profile for MCG-6-30-15, (2) the entire average profile (Nandra et al.\ 
1997) and (3) the broad redshifted part of the average profile. The 
first two options are essentially equivalent so we focus on MCG-6-30-15 
as an extreme example of option 2. This source has perhaps the best 
defined and most studied profile of any Sy I detection. Options 1 and 2
represent the greater challenges because it is certainly easier to 
reproduce a Gaussian profile with a cloud ensemble model.   

We initially assume that the red component is produced by clouds at 
rest at a distance r moving at speed v from the central source. The gravitational 
redshift formula: $$ \lambda / \lambda' = [1-(v/c)\cos \theta] /   
   [\sqrt{(1 - v^2/c^2)( 1 - 2 GM/(c^2 r))}] $$\ (where $\lambda$\ is the
   rest frame wavelength, $\lambda'$\ is the shifted wavelength, $\theta$\
   is the angle between the line of sight and the cloud velocity, M the mass
   of the central black hole) then allows us to convert the observed profile into an 
intensity distribution I(r) contributed by shells at various radii from 
the central source. In the case of 
the redshifted part of the average profile the emission peaks at 
${\rm r}=12 $ \rg\ with a sharp drop towards smaller radii and an 
I(r)$\propto$ 
r$^{-1}$  tail in the opposite direction. The higher redshift of the 
MCG-6-30-15 profile implies even smaller distances from the central source. The 
bulk redshift of the line is a consequence of the fact that the majority 
of the emission arises very close to the black hole. This conclusion is 
true for any kinematical model, including emission from an accretion disk, 
and does not depend on the specific geometric assumptions. 

Unfortunately it is hard to explain why the clouds at only a few gravitational
radii from the black hole would be at rest. The gravitational potential is 
very steep and the radiation pressure is unlikely to be important because  
we are in the sub-Eddington regime. A marginally bound particle passes 
10 \rg\ with a velocity of $0.45$ c during radial infall. With motion 
being dynamically important one must make assumptions about the
geometry and energetics. Our first-order model  considers clouds 
with predominantly radial motion. Angular momentum tends to circularize 
any cloud distribution therefore making it impossible to 
explain the double peaked line profile observed in MCG-6-30-15 (unless a very 
specific planar geometry is assumed as in the case of an accretion disk). 
We constructed models for both radial infall and outflow. As
we show below inflow is clearly preferred by model fits to the  observations. 
This is consistent with the arguments based on energetics. It seems 
very hard to accelerate the matter outwards in a region only a few \rg\
from the black hole. 

The assumptions of the model are as follows: (i) Clouds are in radial bound
orbits defined by their apastron distance R where the clouds would be at rest.
The profile is reproduced from a Monte Carlo simulation of individual cloud
contributions. The distribution of distances r of the clouds is calculated
from the time $dt$ clouds are spending between shells r and ${\rm r} + {\rm
dr}$. (ii) The cloud distribution is limited by the inner  radius ${\rm
r}_{{\rm in}}$  (we assume ${\rm r}_{{\rm in}}=6 $ \rg ) and the outer
radius  ${\rm r}_{{\rm out} } < {\rm R}$  (${\rm r}_{{\rm out}}=15$ \rg\
was assumed throughout). (iii)  The clouds are confined to double cones with a
half-opening angle ${\rm i}_{\rm c}$. (iv)  The inclination of  the line of
sight towards the cone axis is i. (v) Emission in  the comoving frame of the
cloud is isotropic in the optically thin case, and  confined to the inward
looking hemisphere for the case of optically thick  \feka\ emission (both
cases are feasible, see Nandra and George 1994). The  intensity of emission in
the comoving frame does not depend on distance  from the central source. 

Cone models often suffer from a large number of assumptions involving 
geometry and emissivity laws. Our fitting is only illustrative so we  do not
pretend to explore the whole parameter space here. Instead, by  using the
model outlined above, we keep the number of parameters  reasonably low:  only
the apastron distance and the inclination were  fit, with the rest being
assumed or of second order importance. This  is similar to accretion disk
models where the critical parameters are the  inclination and the distance of
its inner edge.

Figure 6 shows the results of our model fits. The thick histogram  shows an
example of the line profile we get for radial inflow of  optically thin clouds
and the thin histogram shows the optically thick  counterpart. The inclination
(${\rm i}=20^\circ$) is intermediate and the cone is  wide (${\rm i}_{\rm c} =
50^\circ$). The value of apastron distance R is important.  Here, as in all
other attempts, we were forced to assume that the velocity  of the clouds
approaches zero at the outer edge of the cloud distribution  which is still
fairly close to the BH (this is consistent with the CLOUDY calculation
presented above). Although on one hand a consequence of this assumption is that 
thesse clouds do not reach large distances from the central source that would
blur the relatively  sharp
observed O~VII edge attributed to the warm absorber (Fabian et al.\ 1994a,
1995), on the other hand it is difficult to explain why the velocity  would
not be higher, i.e. closer to that of marginally bound infalling  clouds. The
accretion disk scenario escapes extreme velocities by invoking nearly face-on
orientation.  In a bi-cone model such a  projection effect  seems unlikely
unless an inclination nearly perpendicular to the cone axis  is assumed. In
this case the contributions from both cones merge in a  single Gaussian-like
redshifted component. It is easy to fit the redshifted  Gaussian component of
the average profile this way, but not the well defined  peak and red shoulder
of the MCG-6-30-15 profile. We do not attempt to  explain the  low radial
velocity of clouds at $\sim 15$\rg\ here. Possible explanations  include a
contribution from an ion torus or from a geometrically thick inner  edge of a
surrounding accretion disk. Mathews (1993) was able to ``bounce''  clouds
around an equilibrium radius (where gravity was balanced by radiation  forces)
about an order of magnitude larger than the one considered here.

It would be satisfying to fit the average profile using the same parameters 
as used for the MCG-6-30-15 fit, varying only the inclination to the line  of
sight. Unfortunately we can not envision a model where this is true, 
accretion disk models included. In the case of a bi-cone the higher
inclination  angle causes the peaks to merge so that the narrow component at
the rest  energy must have a separate origin (an interpretation that we
favor). A  convenient way to fit both components of the average profile with
a  single zone model would involve the assumption of varying optical 
thickness of the emitting clouds. The thin lined histogram in Fig~5 uses the 
same parameters as the MCG-6-30-15 fit, but assuming the optically  thick
regime. The low energy peak is weaker as a consequence of the fact that the
optically thick inward moving clouds on our side of the BH emit most of light
away from us, while we still see the bright faces of the clouds that lie on
the other side and move toward us. The observed average profile lies between
both histogram curves and therefore it is reasonable to conclude  that the
full range of observed \feka\ lines can be explained by different  optical
thickness of the clouds (Nandra and George 1994). 

The model outlined above allows us to distinguish between two kinds of radial
motion, i.e. inflow and outflow. Some of the photons originating in the
far-side cone are intercepted by the BH, so the contribution from the far-side 
cone is weaker. The blue part of the profile gets weaker if inflow is 
considered, while the red part is weaker in the case of outflow. In all  of our
Monte Carlo simulations it was hard to avoid too much flux in the  high energy
wing of the line. The contribution from clouds close to the  black hole moving
towards the observer was acceptably low only for the case  of inflow (though
the high energy flux is still higher than observed,  cf. the thick histogram in
Fig.~5). The distinction between both directions  of motion is even more
obvious in the explanation of the difference between  the MCG-6-30-15 and the
average profile as outlined above. If we deal with outflow it is the high
energy peak that gets weaker when we switch from the optically thin to the 
thick regime, i.e. opposite to the case observed. 

We conclude that a simple double cone model has the potential to explain
both the MCG-6-30-15 and the average \feka\ profile allowing only for a variation of
optical thickness of the emitting clouds. This can be further used to explain
the observed variations in some objects. This is encouraging, still a
physical explanation of the origin of low velocity clouds at $\sim 15$\rg\ is
clearly needed but is beyond the scope of this paper.

\section{Summary of Results \&\ Implications for AGN Models
\label{basta}}

The main results of this investigation are as follows:
\begin{enumerate}

\item The bulk of {\sc Fe} emission near 6.4 keV cannot be directly associated 
with the optical BLR. \feka\ and \hiha\ lines show significantly different 
width and are therefore  not emitted in the same BLR region in radio-quiet 
objects, at least not at the same distance from the central continuum source.

\item A simple illumination model suggests that both \feka\ and \hiha\ could
be emitted by an accretion disk, although the bulk emission for the two lines 
would arise at widely different radii (\feka\ $\simlt$
100 \rg; \hiha $\sim$ 10$^3$ \rg). 

\item While our illuminated disk model is able to predict \hiha\ profile
widths and emitting radii consistent with observational results, the observed and 
model profile shapes show poor agreement. 
The \feka\ emission line data are also inconsistent with a single 
emissivity law. There is poor agreement between disk inclination values
derived for \feka\  and \hiha. Observed \hiha\ asymmetries and shifts 
demand a larger radial drift velocity than compatible with  Keplerian 
disk models. 

\item Evidence is accumulating to suggest that the \feka\ line may be 
composed of (at least) two independent components. 
If the broad component is independent and also shows 
Gaussian symmetry as suggested by the Nandra et al. (1997b) composite spectrum, 
there are no obvious disk models consistent with it. Physically, there are viable
alternatives. The narrow component could
be due to BLR emission with a smaller contribution from the warm absorber,
while the broad component could be redshifted coronal emission. 

\item First-order models involving a biconical distribution of emitting
clouds are able to reproduce the observed \feka\ spectra for AGN. They can
also easily account for the observations that raise serious problems  
for the disk emission scenario.
\end{enumerate}

\section{Acknowledgements}

J. W. S. acknowledges support under NASA AR-05293.01-93A from STScI and
Consiglio Nazionale per le  Ricerche (CNR). P.M.\ acknowledges the hospitality
of UNAM, University of Alabama and the Ljubljana University where parts of 
this work were done. We would like to thank an anonymous referee whose comments greatly
improved the presentation of this paper. 

\section{References}
\REF
Abramowicz, M., Calvani, M., and Nobili, M., 1980, ApJ, 242, 772
\REF
Allen, S., Fabian, A., Idesawa, E., Inoue, H., Kii, T. and Otani, C.,
1997, MNRAS, 286, 765
\REF
Antonucci, R., and Miller, J. S., 1985, ApJ, 297, 621
\REF
Calvani, M., Marziani, P., and Padovani,  P., 1989, in Proceedings of
the VIII Italian Symposium on General Relativity and Gravitation. (Singapore:
World Scientific Co.), 102
\REF
Cassidy I., and Raine, D. J., 1993, MNRAS, 260, 385
\REF
Chen, K., and  Halpern, J. P., 1989, ApJ, 344, 115
\REF
Chen, K., Halpern, J. P. and Filippenko, A. V., 1989, ApJ, 339, 742
\REF
Collin-Souffrin, S. and  Dumont, A. -- M., 1986, A\&A, 166, 13
\REF
Collin-Souffrin, S. and  Dumont, A. -- M., 1990, A\&A, 229, 292
\REF
Cunningham, Chr., 1976, ApJ, 208, 534
\REF
Dabrowski, Y., Fabian, A. C., Iwasawa, K., Lasenby, A. N., \&\ Reynolds, C.
S., 1997, MNRAS, 288, L11
\REF
Done, C. et al.\ 1996, ApJ, 463, L63
\REF
Dumont, A. -- M., and Collin-Souffrin, S., 1990a, A\&AS, 83, 71
\REF
Dumont, A. -- M., and  Collin-Souffrin, S., 1990b, A\&A, 229, 313
\REF
Dultzin-Hacyan, D., 1997, RMxAA Serie de Conferencias, 6, 132
\REF
Elvis, M. et al., 1994, ApJ, 436, L55
\REF
Eracleous, M., and Halpern, J. P., 1994, ApJS, 90, 1
\REF
Eracleous, M., et al., 1996, ApJ, 459, 89
\REF
Fabian et al., 1994a, PASJ, 46, L59
\REF
Fabian, A. et al., 1994b, ApJ, 436, L51
\REF
Fabian, A. C., Nandra, K., Reynolds, C. S., Brandt, W. N., Otani, C.,
\&\ Tanaka, Y. 1995, MNRAS, 277, L11
\REF
Fanton, C., Calvani, M., de Felice, F., and  \v Cade\v z, A. 1997, PASJ, 49, 159
\REF
Ferland, G. J., and  Rees, M. J., 1988, ApJ, 332, 141
\REF
Ferland, G. J., 1996, Hazy: a brief introduction to CLOUDY, University of
Kentucky, Dept.\ Phys.\ Astron., internal report
\REF
Filippenko, A. V., and  Sargent, W. L. W., 1988, ApJ, 324, 134
\REF
Frank, J., King, R., and  Raine, D., 1992, Accretion Power in
Astrophysics (2nd edition), (Cambridge: Cambridge University Press)
\REF
Fukazawa, Y. et al., 1994, PASJ, 46, L141
\REF
Gaskell, C. M., 1982, ApJ, 263, 79
\REF
Gaskell, C. M., 1996, ApJ, 464, L107 
\REF
George, I. M., \&\ Fabian, A. C., 1991, MNRAS, 249, 352 
\REF 
George, I. et al., 1995, ApJ, 438, L67
\REF
Ghisellini, G., Haardt, F. and Matt, G., 1994, MNRAS, 267, 743 
\REF
Gonzalez--Delgado, M., and Perez, E., 1996, MNRAS, 278, 737
\REF
Guainazzi, M., Matsuoka, M., Piro, L., Mihara, T., and Yamauchi,, M., 1994,
ApJ, 436,  L35
\REF
Guainazzi, M., Mihara, T., Otani, C., and  Matsuota, M., 1996, PASJ, 48, 781
\REF
Guilbert, P. W., and Rees, M. J., 1988, MNRAS, 233, 475
\REF
Halpern, J., et al., 1997, ASCA Cherry Blossom Conference 
\REF
Hayashi, I. et al., 1996, PASJ, 48, 219
\REF
Ho, L. C., Filippenko, A. V., and Sargent, W. L. W., 1995, ApJS, 98, 477
\REF
Ishisaki, Y., et al., 1996, PASJ, 48, 237
\REF 
Iwasawa, K., and Taniguchi, Y, 1993, ApJ, 413, L15
\REF 
Iwasawa, K., et al., 1994, PASJ, 46, L167
\REF
Iwasawa, K., et al., 1996a, MNRAS, 279, 837
\REF
Iwasawa, K., et al., 1996b, MNRAS, 282, 1038
\REF
Laor, A., Fiore, F., Elvis, M., Wilkes, B., \&\ McDowell, J., 1997, ApJ, 477, 93
\REF
Lightman, A. P., \&\ White, 1988, ApJ, 335, 57
\REF
Krolik, J. H., Madau, P., \&\ \.{Z}ycki, P. T., 1994, ApJ, 420, L57
\REF
Madau, P., 1988, ApJ, 327, 116
\REF
Makishima, K. et al., 1994, PASJ, 46, L77
\REF
Martocchia, A., and  Matt, G., 1996, MNRAS, 282, L53 
\REF
Marziani, P., Calvani, M., and  Sulentic, J. W., 1992, ApJ, 393, 658
\REF
Marziani, P., Sulentic, J. W., Dultzin-Hacyan,, D., Calvani, M.,
\&\ Moles, M., 1996, ApJS, 104, 37
\REF
Mathews, W. G., 1993, ApJ, 412, L17
\REF
Mathews, W. and Ferland, G., 1987, ApJ, 323, 456 (MF87)
\REF
Mathews, W. G., \&\ Capriotti, E. R., 1984, in Astrophysics of Active Galaxies
and Quasi-stellar Objects, J. S. Miller (Ed.), (Mill Valley: University Science
Books), p. 184 
\REF
Matt, G., Perola, G. C. and Stella, L., 1992, A\&A, 257, 63
\REF
Matt, G., Fabian, A. C., and   Ross, R.R.,  1996, MNRAS, 278, 111 
\REF
Mihara, T. et al., 1994, PASJ, 46, L137
\REF
Morris, S. L.,  Ward, M. J., 1988, MNRAS, 230, 639
\REF
Morris, S. L.,  Ward, M. J., 1989, ApJ, 340, 713
\REF
Murray, N., \&\ Chiang, J., 1997, ApJ, 474, 91
\REF
Mushotzky, R. F., 1997, RMxAC, 6, 213
\REF
Mushotzky, R. F., Done, C., and Pounds, K. A., 1993, ARA\&A, 31, 717
\REF
Mushotzky, R. F., Fabian, A. C., Iwasawa, K., Kunieda, H., Matsuoka, 
M., Nandra, K., and  Tanaka, Y., 1995, MNRAS, 272, L9
\REF
Nandra, K. and George, I., 1994, MNRAS, 267, 974
\REF
Nandra, K., George, I. M., Turner, T. J. and Fukazawa, Y., 
1996,  ApJ, 464, 165 
\REF
Nandra, K., George, I. M., Mushotzky, R. F., Turner, T. J., \&\ Yaqoob, 1997a,
ApJ, 476, 70
\REF
Nandra, K., George, I. M., Mushotzky, R. F., Turner, T. J. and Yaqoob, T., 
1997b, ApJ, 477, 602 
\REF
Nandra, K., George, I. M., Mushotzky, R. F., Turner, T. J. and Yaqoob, T., 
1997c, ApJ, 488, L91 
\REF
Nelder, J. A., and Mead, R., 1965, Computer Journal, 7, 308
\REF
Novikov, I., and Thorne, K. S., 1973, In: Black Holes, Eds. De Witt, B. S. 
and De Witt, C., (Gordon \&\ Breach: New York)  
\REF
Osterbrock, D. E., and Pogge, R. W., 1985, ApJ, 297, 166
\REF
Osterbrock, D. E., and Shuder, J. M., 1982, ApJS, 49, 149
\REF
Padovani, P., 1989, A\&A, 209, 27
\REF
Pounds, K., et al., MNRAS, 1994, 267, 193
\REF
Pounds, K. et al., 1995, MNRAS, 277, L5
\REF 
Ptak, A. et al., 1994, ApJ, 436, L31
\REF 
Ptak, A. et al., 1996, ApJ, 495, 542
\REF
Puchnarewicz, E. M., Mason, K. O., Siemiginowska, A., and  Pounds, K. A., 1995,
MNRAS, 276, 20
\REF
Rafanelli, P., and  Schulz, H., 1983, A\&A, 117, 109
\REF
Reynolds, C., 1997, MNRAS, 286, 513
\REF
Reynolds, C. S., and  Fabian, A. C., 1995, MNRAS, 273, 1167 
\REF
Reynolds, C. S. et al., 1995, MNRAS, 276, 1311 
\REF
Rokaki, E., Collin-Souffrin, S., and  Magnan, C., 1993, A\&A, 272, 8 
\REF
Romano, P., Zwitter, T., Calvani, M., Sulentic J. W., 1996, MNRAS, 279,
165
\REF
Shakura, S., \&\ Sunyaev, R., 1973, A\&A, 24, 337
\REF
Shuder, J. M., 1981, AJ, 86, 1595
\REF
Sivron, R. and Tsuruta, S., 1993, ApJ, 402, 420
\REF
Smith, D. and Done, C., 1996, MNRAS, 280, 355
\REF
Sulentic, J. W., 1989, ApJ, 343, 54
\REF
Sulentic, J. W., Calvani, M., Marziani, P., \&\ Zheng, W., 1990, ApJ,
355, L15
\REF
Sulentic, J. W., Marziani, P., Zwitter, T. and Calvani, M. 1995a, 
ApJ, 438, L1
\REF
Sulentic, J. W., Marziani, P., Dultzin-Hacyan, D., Calvani, M,
and  Moles, M., 1995b, ApJ, 445, L85
\REF
Tanaka, Y., Holt, S. S., and  Inoue, H., 1994, PASJ, 46, L37
\REF
Tanaka, Y., et al.,  1995, Nature, 375, 659
\REF
Thorne, K., 1974, ApJ, 191, 507
\REF
Turner, T. J., et al., 1996, ApJ, 463, 134
\REF
Turner, T. J. George, I. M., Nandra, K., Mushotzky, R. F., 1997, ApJ, 488, 164
\REF
Ueno, S. et al., 1994, PASJ, 46, L31
\REF
Yaqoob, T. et al., 1994a, PASJ, 46, L49
\REF
Yaqoob, T. et al., 1994b, PASJ, 46, L173
\REF
Yaqoob, T., Edelson, R., Weaver, K. A., Warwick, R. S., Mushotzky, R.
F., Serlemitsos, P. J., and Holt, S. S., 1995, ApJ, 453, L81
\REF
Weaver, K. A., Nousek, J., Yaqoob, T., Hayashida, K., and Murakami, S.,
1995, ApJ, 451, 147
\REF
Weaver, K. A., Nousek, J., Yaqoob, T., Mushotzky, R. F., 
Makino, \&\ Otani, C.,  1996, ApJ, 458, 160
\REF
Weaver, K., Yaqoob, T., Mushotzky, R. F., Nousek, J., Hayashi, I. and Koyama, 
K., 1997, ApJ, 474, 675
\REF
Zheng, W., Binette, \&\ Sulentic, J. W., 1990, ApJ, 365, 115
\REF
Zheng, W., Veilleux, S. and Grandi S., 1991, ApJ, 418, 196
\REF
\.{Z}ycki, P. T., and  Czerny, N., 1994, MNRAS, 266, 653

\newpage

\begin{figure}
\caption[]{Upper: Spectrum of Arp102B in the region of \hiha. Taken  from a 35
minute exposure at the Asiago Observatory 1.82 m. with  a dispersion of 120
\AA/mm (March 26, 1989 UT=0254). Lower:  Same spectrum  degraded to the
sampling rate of an ASCA observation of  \feka\ (after narrow lines
subtraction). Solid line: disk model profile fitted to the \ha\ profile by Chen
\&\ Halpern (1989); dot-dashed line gaussian profile. }
\end{figure}

\begin{figure}
\caption[]{ FHWM of \feka\ versus FWHM of \hiha, in \kms. 
Filled circles: SIS ASCA data; filled squares: BLRG; circled filled circles: 
best candidate disk emitters; the star is NGC 1068 (broad \ha\ component visible 
in polarized light). The dot-dashed line is the line of equal \feka\ and \hiha\ FWHM. }
\end{figure} 

\begin{figure}
\caption[]{Plots of \hiha and \feka\ surface emissivity, in units of ergs
s$^{-1}$ cm$^{-2}$, derived from our illumination models as a function of disk
radius. Full line: ``cold'' \feka\ (iron less than 16 times ionized)
emissivity; dashed line: ``hot'' \feka\ emissivity; dotted line: \hiha\
emissivity.  See text and Table 4 for model parameters. Labels identify the
model as in Table 4. ``Hot'' iron emission has been plotted for model (2) only
for clarity. \ha\ surface emissivity has been plotted for r $\simgt $ 100 \rg.}
\end{figure} 

\begin{figure}
\caption[]{\feka\ line profile for MCG-6-30-15 as obtained by Tanaka et al.\ 
1995. The solid line is our best  Kerr disk profile  model fit with 
${\rm r}_{{\rm in}}\approx  1.25 $\rg, ${\rm r}_{{\rm out}} \approx 20 $\rg, 
i = 30$^\circ$ and power-law emissivity with 
spectral index p = $-1.8$.}
\end{figure}

\begin{figure} 
\caption[]{ Optical \ha\ line profiles of 15 AGN
considered in this paper, after subtraction of the underlying  continuum.
Vertical scale is specific flux (ergs s$^{-1}$ cm$^{-2}$ \AA$^{-1}$ $\times$
10$^{15}$); horizontal  scale is wavelength in \AA. The thick lines show model
profiles computed following Chen \&\ Halpern (1989).
Dotted lines indicate the pure broad line (narrow lines substracted).
 See Table 6 for
model parameters. }
\end{figure}

\begin{figure}
\caption[]{
Observations and theoretical models of the \feka\ line. {\it
Points:} observed data for MCG-6-30-15, {\it dashed curve:} accretion disk fit
(Tanaka et al.\ 1995), {\it solid line:} observed averaged profile 
for 16 sources (Nandra et al.\ 1997). {\bf (a)} {\it thick histogram:} 
double-cone optically thin radial inflow of $10^4$ clouds (${\rm i} = 
20^\circ$, ${\rm i_c} = 50^\circ$,  ${\rm r} = 16$\rg:  see text), 
{\it thin histogram:} same for optically thick inflow. }
\end{figure}
\end{document}


\begin{small}
\begin{deluxetable}{llrrlllr}
\small
\tablewidth{0pc}
\tablenum{1}
\tablecaption{Table of  ASCA FeK line detections}
\tablehead{
\colhead{IAU Code} & \colhead{Name} & \colhead{Type} & \colhead{Z} &
\colhead{Fcent} & \colhead{Fsig} &\colhead{F(EW)} &\colhead{Ref}\\
\colhead{(1)} &
\colhead{(2)} &
\colhead{(3)} &
\colhead{(4)} &
\colhead{(5)} &
\colhead{(6)} &
\colhead{(7)} &
\colhead{(8)} 
}
\startdata
%

J0006+202 & Mark 335  & Sy1   & 0.025 & 6.4    & \nodata  & 110     & 12 \nl
J0017+816 & S5 0014+813  & QSO   & 3.387 & 6.4    &  nar.    & $<$120  & 31 \nl
J0123$-$588 & Fairall 9 & Sy1   & 0.046 & 6.4    & 0.35     & 300-400 & 12 \nl
J0228+313 & Mark 1040 & Sy1   & 0.016 & 6.4    & 0.33     & 550     & 18 \nl
J0242$-$000 & NGC 1068  & Sy2   & 0.003 & 6.4/.7 & 0.1/.3   & 11-2600 & 21 \nl
J0413+112 & 3C109    &  RG   & 0.305 & 6.6    & 0.65     & 300     & 23 \nl
J0433+053 & 3C120    & RG    & 0.033 & 6.4    & 0.7      & 330     & 12 \nl
J0552$-$074 & NGC 2110  & Sy2   & 0.007 & 6.4    & 0.1      & 124     & 22 \nl
J0558$-$383 & EXO 0556.3-3820  & Sy1   & 0.034 & 6.3    & 0.5      & 277     &  9 \nl
J0615+710 & Mark 3    & Sy2   & 0.014 &   *    &   *      &  *      & 20 \nl
J0913+409 & IRAS 09104+4109 & QSO   & 0.442 & 6.6    & 0.07     & 450     & 29 \nl
J0934+273 & MCG -5-23 & Sy3    & 0.044 &   *    &    *     &  *      & 28 \nl
J0945$-$143 & NGC 2992  & Sy1.9 & 0.008 & 6.4    & 0.03     & 514     & 16 \nl
J0955+690 & NGC 3031  & Sy1   & 0.000 & 6.6/.9 & 0.2/.4   & 170-220 &  8 \nl
J1016+734 & NGC 3147  & Sy2   & 0.010 & 6.3/.6 & \nodata  & 370-700 & 19 \nl
J1023+198 & NGC 3227  & Sy1   & 0.003 & 6.4v   & 0.1/.3   & 250     & 13 \nl
J1034+396 & RE J1034+393   & Sy1   & 0.042 & 6.8    & \nodata  & 890     & 17 \nl
J1106+725 & NGC 3516  & Sy1   & 0.009 & 6.1    & 0.6      & 340     & 12 \nl
J1119+213 & PG 1116+215   & QSO   & 0.177 & 6.9    & 0.4      & 280     &  4 \nl
J1139$-$377 & NGC 3783  & Sy1   & 0.009 & 6.3    & 0.7      & 360-510 & 12, 15 \nl
J1203+445 & NGC 4051  & Sy1   & 0.002 & 6.0    & 0.1      & 160     &  10 \nl
J1210+394 & NGC 4151  & Sy1.5 & 0.003 & 6.4    & 0.7-0.9  & 300-400 &   2 \nl
J1214+140 & PG 1211+143   & QSO   & 0.085 & 6.6    & 0.48     & 198     & 24 \nl
J1218+298 & Mark 766  & Sy1   & 0.012 & 6.0    & 0.7      & 550     & 12 \nl
J1218+473 & NGC 4258  & Sy3   & 0.002 & 6.5    & 0.2      & 250     & 27 \nl
J1229+020 & 3C 273    & QSO   & 0.158 & 6.4    & 0.1      &  19     & 26 \nl
J1239$-$053 & NGC 4593  & Sy1   & 0.009 & 6.35   & \nodata  &  90     &  12 \nl
J1305$-$495 & NGC 4945  &  Sy   & 0.002 &   *    &   *      & *       & 30 \nl
J1335$-$342 & MCG -6-30-15 & Sy1   & 0.008 & 6.4v   & 0.2-0.7  & 210-330 &  1 \nl
J1349$-$303 & IC 4329A  & Sy1   & 0.016 & 6.3    & 0.1-0.5  & 80-230  &  3 \nl
J1417+251 & NGC 5548  & Sy1   & 0.017 & 6.1    & 0.5      & 154     &  2 \nl
J1504+104 & Mark 841  & Sy1   & 0.036 & 6.4    & 0.63     & 580     & 12 \nl
J1800+666 & NGC 6552  & Sy2   & 0.026 & 6.4    & \nodata  & 300-900 & 25 \nl
J1836$-$594 & IRAS 1832-5926 & Sy2   & 0.019 & 6.8    & 0.85     & 580     & 6  \nl
J1842+797 & 3C 390.3  & RG    & 0.057 & 6.3    & 0.15     & 190     & 5  \nl
J1942$-$103 & NGC 6814  & Sy1   & 0.006 & 6.3    & 0.6      & 1100    & 12 \nl
J2044$-$107 & Mark 509  & Sy1   & 0.035 & 6.4    & 0.47     & 200-300 & 12 \nl
J2235$-$260 & NGC 7314  & Sy1   & 0.005 & 6.2    & 0.87     & 458     & 11 \nl
J2303+088 & NGC 7469  & Sy1   & 0.017 & 6.3/.9 & 0.15     & 120-200 & 14 \nl
J2304$-$086 & MCG -2-58 & Sy1   & 0.047 & 6.2    & 0.35     & 340     & 7  \nl
\tablecomments{
Col.\ (1) - abbreviated (2000) position taken from the 7th edition of
the Veron \& Veron catalog; Col.\ (2) - a common name for the source as
used in the ASCA-related literature; Col.\  (3) - AGN type; Col.\ (4) -
optically determined redshift; Col.\ (5), (6), and (7) - the derived \feka\
line peak centroid energy, profile Gaussian width (keV) and EW (eV),
respectively; Col.\  (8) - reference(s) to the source of the \feka\
parameters given in the three previous columns.  
*: objects with clear multicomponent FeK line profiles, see
references for details.}
\tablerefs{
1: Tanaka et al., 1995;
2: Yaqoob et al., 1995;
3: Mushotzky et al., 1995;
4: Nandra et al., 1997;
5: Eracleous et al., 1996;
6: Iwasawa et al., 1996a;
7: Weaver et al., 1995;
8: Ishisaki et al., 1996;
9: Turner et al., 1996;
10: Mihara et al., 1994;
11: Pounds et al., 1994;
12: Nandra et al., 1997;
13: Ptak et al., 1994;
14: Guainazzi et al., 1996;
15: George et al., 1995;
16: Weaver et al., 1996;
17: Pounds et al., 1995;
18: Reynolds et al., 1995;
19: Ptak et al., 1996;
20: Iwasawa et al., 1994;
21: Ueno et al., 1994;
22: Hayashi et al., 1996;
23: Allen et al., 1997;
24: Yaqoob et al., 1994a;
25: Fukuzawa et al., 1994;
26: Yaqoob et al., 1994b;
27: Makishima et al., 1994;
28: Weaver  et al., 1997; 
29: Fabian et al., 1994;
30: Done et al., 1996, ApJ;
31: Elvis et al., 1994.
}
\enddata
\end{deluxetable}

\newpage

\begin{deluxetable}{llrrcclrrc}
\tablewidth{0pt}
\tablenum{2}
\tablecaption{Journal of H$\alpha$ Line Profile Observations}
\tablehead{ 
\colhead{IAU Code} & \colhead{Name} &\colhead{Date} & \colhead{UT} 
&\colhead{ET} & \colhead{Obs.} & \colhead{Tel.} & \colhead{Spectr.} &\colhead{Grat.} & \colhead{Res. FWHM} \\
\colhead{(1)} & \colhead{(2)} & \colhead{(3)} & \colhead{(4)} & \colhead{(5)} &\colhead{(6)}
&\colhead{(7)}&\colhead{(8)}& \colhead{(9)} & \colhead{(10)} } 
\startdata
\\
J1106+725   & NGC 3516     & Mar. 24, 1989 & 21:53 & 2400 & Asiago & 1.82m & B\&C &  600 gr/mm  & 4 \nl
J1210+394   & NGC 4151     & Mar. 24, 1989 & 00:42 & 1800 & Asiago & 1.82m & B\&C &  600 gr/mm  & 4 \nl
J1239$-$053 & NGC 4593     & Apr.  2, 1990 & 06:03 & 3600 & ESO    & 1.52m & B\&C & 1200 gr/mm & 2     \nl
J1335$-$342 & MCG -06-30-15 & Feb.  3, 1995 & 11:39 & 2400 & SPM    & 2.12m & B\&C & 1200 gr/mm & 2   \nl
J1504+104   & Mark 841     & May  21, 1993 & 04:13 & 3300 & ESO    & 1.52m & B\&C & 1200 gr/mm  & 2   \nl
J1942$-$103 & NGC 6814     & May  23, 1993 & 06:37 & 1800 & ESO    & 1.52m & B\&C &  600 gr/mm  &  4  \nl
J2044$-$107 & Mark 509     & May, 25, 1993 & 09:33 & 3600 & ESO    & 1.52m & B\&C  &   600 gr/mm  & 4 \nl
J2303+088   & NGC 7469     & Aug. 20, 1990 & 10:28 & 1200 & SPM    & 2.12m & B\&C & 600 gr/mm  & 4      \nl
J2304$-$086 & Mark 926     & Sept.26, 1989 & 03:12 & 3000 & ESO    & 1.52m & B\&C & 1200 gr/mm & 2      \nl
\tablecomments{
Col.\ (1) - abbreviated IAU designation; Col.\ (2)
- other name;  Col.\ (3) - date of observation; Col.\  (4) - universal  time of
observation; Col.\ (5) -  exposure time  in seconds; Col.\  (6) - observatory;
Col.\ (7) -  telescope; Col.\  (8) -  spectrograph; Col.\  (9) -  grating;
Col.\ (10) - resolution FWHM (in \AA).}
\enddata
\end{deluxetable}
\newpage

\begin{deluxetable}{llrcrcclr}
\small
\tablewidth{0pt}
\tablenum{3}
\tablecaption{Broad  H$\alpha$ Line Profiles}
\tablehead{
\colhead{IAU Code} & \colhead{Name} & \colhead{S. T.} &
\colhead{EW} &
\colhead{FWHM} & \colhead{$\Delta$(FWHM)} &\colhead{$\rm v_r$(1/2)} & \colhead{$\Delta\rm v_r$} 
&\colhead{Ref.} \\
\colhead{(1)} & \colhead{(2)} & \colhead{(3)} & \colhead{(4)} & \colhead{(5)} &\colhead{(6)}
&\colhead{(7)}&\colhead{(8)} &\colhead{(9)}} 
\startdata
\\
J0006+202*               &  Mark 335     & Sy1 &   490    &  1600   & 300   & 300 & 100 &  1,2\nl
J0123$-$588*$^\dagger$   &  Fairall 9     & Sy1 &   420    &  4550   & 500   & 410 & 200 &  3\nl
J0228+313                & Mark 1040      & Sy1 &   305       &  2200   & 300   &  0    &  200 & 4,5\nl
J0242$-$000$^\clubsuit$  & NGC 1068       & Sy2 &   \nodata          &  4100    & 1000 & \nodata &  \nodata &  6\nl
J0433+053*$^\dagger$       & 3C 120       & RG  &   470                 &   2400  & 300  & 110 & 100 & 7 \nl
J0558$-$383            & EXO 0556.3-3820  & Sy1 &   760                    &    4250     & 400       & -160 & 100 & 8\nl
J0913+409              & IRAS 0910        & QSO &   \nodata & \nodata & \nodata & \nodata & \nodata & \nodata\nl
J0955+690              &NGC 3031           & Sy1&   26      &     3400    &   350    &   -140    &  200   & 9,10  \nl
J1023+198*$^\dagger$   &NGC 3227          & Sy1 &   285     &     4000    &   400    &       70    &   100  &   7 \nl
J1034+396             & ZW 212.025      & Sy1  &   200     &     1800    &   200    &         0    &   200   & 11 \nl
J1106+725*$^\dagger$       & NGC 3516   & Sy1   &   150     &     5300    &   500     &   -270   &   100   &    12\nl
J1119+213$^\dagger$         & PG 1116+215    & QSO &   340     &     2750    &   400    &        60  &   100   &    7   \nl
J1139$-$377*$^\dagger$    & NGC 3783     & Sy1  &   440     &     3050    &   300    &        40  &   100   &    3  \nl
J1203+445*                &  NGC 4051  &   Sy1  & \nodata   &     1000    &   200    &     100   &   100    &   5 \nl
J1210+394*                & NGC 4151   & Sy1.5  &   240     &    6400    &  600    &    -350   &   100    &  12   \nl
J1214+140$^\dagger$       &PG 1211+143 &QSO     &   390     &     1700  &  300    &      -30   &   100    &  7  \nl
J1218+298*               &Mark 766     &Sy1     &  $^\spadesuit$195    &     1600  &  200    &          0   &   200     &  13,14  \nl
J1229+020                & 3C 273       &QSO    &   613     &     3400  & 400     &     100   &   100     &  15,2\nl
J1239$-$053*$^\dagger$    & NGC 4593  & Sy1     &   250        & 3400  & 400     &    -180   &    70      &  12 \nl
J1335$-$342*$^\dagger$   & MCG -6-30-15 & Sy1   &   160        & 2200  &  200    &        50    &  100     & 12 \nl
J1349$-$303*$^\dagger$   & IC 4329A    & Sy1    &   515        & 5000  & 400    &     330    &   150     &  16\nl
J1417+251*$^\dagger$     & NGC 5548    & Sy1    &   540        & 5450   & 300   &    -590   &    200      & 3 \nl
J1504+104*$^\dagger$    & Mark 841     & Sy1    &   470        & 5000  & 400    &   -220    &   100      & 12\nl
J1842+797               & 3C 390.3     & RG     &   450        & 12000 & 1000 &     300   &    150     &  12,7 \nl
J1942$-$103*$^\dagger$ & NGC 6814       & Sy1   &    50        & 5600  & 1000 &     360  &     200     & 12 \nl
J2044$-$107*$^\dagger$  & Mark 509     &  Sy1   &   550        & 3200  &  400  &   -200  &     100     &  12 \nl
J2303+088*$^\dagger$      & NGC 7469    & Sy1   &   380        & 2950  & 300   &    320   &    100     & 12\nl
J2304$-$086*$^\dagger$ & Mark 926      &  Sy1  &   800        & 7650     & 1500  & 
 790& 300 & 12 \nl
\tablecomments{
Col.\ (1) - IAU code name; Col.\  (2) - 
source name; Col.\  (3) - Seyfert type; Col.\  (4) - EW in \AA; Col.\  (5) - FWHM in
km/s;   Col.\  (6) - uncertainty at  2$\times \sigma$\ confidence level; 
 Col.\ (7) -  centroid line shift (at half maximum) in  \kms; Col.\  (8) -  line centroid
uncertainty; Col.\ (9) - reference code. 
*: objects studied by Nandra et al.\ 1997; $\dagger$: H$\alpha$\ profile 
shown in Fig. 2; $^\clubsuit$: 
broad component revealed in polarized light; $^\spadesuit$:
 estimated as EW(H$\beta$) $\times$ Balmer decrement.}
\tablerefs{
1: Shuder, 1981;
2: Cassidy \&\ Raine, 1993;
3: Marziani et al., 1996; 
4: Rafanelli \&\ Schulz, 1983;
5: Osterbrock \&\ Shuder, 1982;
6: Antonucci \&\ Miller, 1985;
7: G. Stirpe, private communication \&\ unpublished thesis work; 
8: Morris \&\ Ward, 1989;
9: Filippenko \&\ Sargent, 1988;
10: Ho, Filippenko \&\ Sargent, 1995;
11: Puchnarewicz et al., 1995
12: this work;
13: Osterbrock \&\ Pogge, 1985;
14: Gonzalez-Delgado \&\ Perez, 1996;
15: Morris \&\ Ward, 1988;
16: Marziani, Calvani, \&\ Sulentic, 1992.}
\enddata
\end{deluxetable}

\newpage

\begin{deluxetable}{lcccc}
\tablenum{4}
\small\small
\tablewidth{0pt}
\tablecaption{Explorative Photoionization Computations}
\tablehead{
\colhead{Source} & \multicolumn{2}{c}{$\log$ L(\feka)} &
\colhead{{\sc Fe} Ionization} & \colhead{$\log_{10}$(\hiha)}  
\\ \cline{2-3}
\colhead{} & 
\colhead{A({\sc Fe})$_\odot$} & 
\colhead{10$\rm \times$ A({\sc Fe})$_\odot$} &
\colhead{} & \colhead{} \\
\colhead{(1)} &
\colhead{(2)} &
\colhead{(3)} &
\colhead{(4)} &
\colhead{(5)} }
\startdata  
BLR 		&   39.65         &	40.50	&  II  &  41.5 		\nl
Warm Absorber	&   39.22         &	40.24	&  IX  &  40.0 		\nl
Coronal gas	&   $\simgt$35.7  &	$\simgt$ 38.2	&  XXVII    & 32.4 \nl
Dusty torus     & $\simlt$ 38.85  &	$\simlt$ 39.28	&  IV  & 
 \nodata\tablenotemark{a}   \nl
\tablecomments{
Col.\ (1) - possible origin for \feka\ line emission; 
Col.\ (2) and (3) - the \feka\ luminosity computed in the case of 
gas with
solar chemical abundances and of a factor 10 iron overabundance; 
Col.\ (4) - the dominant
ionization stage of iron;  Col.\  (5) -  the expected \hiha\
luminosity.} 
\enddata
\tablenotetext{a}{L({\sc Hi~H}$\alpha$) cannot be reliably predicted by an optically thin torus
model.}
\end{deluxetable}
\newpage

\begin{deluxetable}{cclcccccc}
\tablenum{5}
\small
\tablewidth{0pt}
\tablecaption{Disk Model Results for MCG -06-30-15}
\tablehead{\colhead{Model} &
\colhead{r$\rm _{halo,in}$} & 
\colhead{r$\rm _{halo,out}$} & \colhead{$\rm \tau_{es}$} &
\colhead{$\rm \tau_{es}$\ Halo} & Continuum & 
\colhead{$\log_{10}$L(\feka)} &
\colhead{{\sc Feii}$\rm _{Hot}$/{\sc Feii}$ \rm _{cold}$} &
 \colhead{$\log_{10}$L(\hiha)} \\ 
\colhead{(1)} & \colhead{(2)} &
\colhead{(3)} & \colhead{(4)} & \colhead{(5)} & \colhead{(6)}
& \colhead{(7)} & \colhead{(8)} & \colhead{(9)}}
\startdata  
(1) & 1. & 100. & 0.25& 0.25  & MF97	&  41.86	  & $\simlt 0.3$  & 41.62  \nl	
(2) & 1. & 100.  & 0.25&  --  & MF97      &  41.36	  &	0.1	  & 40.63  \nl
(3) & 1. & 100. & 0.25& 0.25  & power-law &  40.56	  &  negligible	  & 39.37  
\nl
\tablecomments{
Col.\  (1) - identification number for each model;  Col.\ (2) (3),
(4) and  (5) - the inner and outer radii,  the electron scattering optical depth
  for
the inner sphere and the outer halo; Col.\ (6) - the shape of
the continuum;  Col.\  (7), (8) and (9) -
 the logarithms of the \feka\ luminosity from cold iron emission, 
the ratio L(\feka)$\rm_{Hot}$/L(\feka)$\rm _{Cold}$, and  the $\rm \log(L($\hiha))
respectively.}  

\enddata
\end{deluxetable}
\newpage

\begin{deluxetable}{llcccccc}
\small
\tablewidth{0pt}
\tablenum{6}
\tablecaption{Broad  H$\alpha$ Disk Model Fitting Results}
\tablehead{
\colhead{IAU Code} & \colhead{Name} & \colhead{R$_{\rm out}$}  &
\colhead{i$_{\rm opt}$[1]} & \colhead{$\chi^2$[1]} 
&
\colhead{i$_{\rm opt}$[2]} & \colhead{$\chi^2$[2]} 
&\colhead{i$_{\rm Fe}$}
\\
\colhead{(1)} & \colhead{(2)} & \colhead{(3)} & \colhead{(4)} & \colhead{(5)} 
& \colhead{(6)} & \colhead{(7)} & \colhead{(8)} }
\startdata
\\
J0123$-$588 & Fairall 9   & 1.0$\times$10$^5$  & 36$^\circ$ & 6.16 & 15$^\circ$ & 0.90  & 89$^\circ$\nl
J0433+053   & 3C 120      & 1.9$\times$10$^5$  & 34$^\circ$ & 1.00 &  9$^\circ$ & 1.58  & 88$^\circ$\nl
J1023+198   & NGC 3227    & 1.0$\times$10$^5$  & 36$^\circ$ & 4.88 & 12$^\circ$ & 0.81  & 21$^\circ$\nl
J1106+725   & NGC 3516    & 1.5$\times$10$^5$  & 36$^\circ$ & 9.96 & 20$^\circ$ & 1.63  & 26$^\circ$\nl
J1139$-$377 & NGC 3783    & 2.0$\times$10$^5$  & 50$^\circ$ & 1.00 & 12$^\circ$ & 2.25  & 26$^\circ$\nl
J1210+394   & NGC 4151    & 1.5$\times$10$^5$  & 50$^\circ$ & 9.44 & 25$^\circ$ & 1.38	& 24$^\circ$\nl
J1239$-$053 & NGC 4593    & 1.0$\times$10$^5$  & 38$^\circ$ & 1.36 & 14$^\circ$& 0.76  & 0$^\circ$\nl
J1335$-$342 & MCG-6-30-15 & 1.0$\times$10$^6$  & 13$^\circ$ & 0.46 &  9$^\circ$ & 2.95  & 34$^\circ$\nl
J1349$-$303 & IC 4329A    & 1.0$\times$10$^5$  & 36$^\circ$ & 13.6 & 12$^\circ$ & 5.24	& 10$^\circ$\nl
J1417+251   & NGC 5548    & 1.0$\times$10$^5$  & 37$^\circ$ &  7.64 & 18$^\circ$ & 0.25	& 10$^\circ$\nl
J1504+104   & Mark 841    & 1.0$\times$10$^5$  & 38$^\circ$ & 4.36 & 17$^\circ$ & 0.83	& 26$^\circ$\nl
J1942$-$103 & NGC 6814    & 1.0$\times$10$^5$  & 38$^\circ$ & 0.67 & 19$^\circ$ & 0.1	&  \dots\nl
J2044$-$107 & Mark 509    & 1.0$\times$10$^5$  & 35$^\circ$ & 4.64 & 12$^\circ$ & 7.08  & 89$^\circ$\nl
J2303+088   & NGC 7469    & 9.0$\times$10$^5$  & 50$^\circ$ & 1.04 & 10$^\circ$ & 0.97  & 20$^\circ$\nl
J2304$-$086 & Mark 926    & 1.0$\times$10$^5$  & 38$^\circ$ & 35.0 & 26$^\circ$ & 2.77  & 26$^\circ$\nl
\tablecomments{
All objects studied by Nandra et al.\ 1997 and 
with H$\alpha$\ profile shown in Fig.\ 5.
Col.\ (1) and (2) - IAU code and common name; Col.\  (3) - disk outer radius; 
Col.\  (4) - \hiha\ disk inclination from our model 1 fits; 
Col.\ (5) - the reduced $\chi^2$; Col.\  (6) - \hiha\ disk inclination from our model 2 fits; 
Col.\ (7) - the reduced $\chi^2$; Col.\ (8)  - \feka\  disk inclination; 
data taken from Nandra et 
al.\ (1997) for Kerr geometry with reflection: in cases where 
multiple observations were presented, we chose the inclination
with the smallest uncertainty.}
\enddata
\end{deluxetable}
\end{small}